# Lattice-Constant and Band-Gap Tuning in Wurtzite and Zincblende BInGaN Alloys

Kevin Greenman,[1] Logan Williams,[2] and Emmanouil Kioupakis[2,a)]

[1]*Department of Chemical Engineering, University of Michigan, Ann Arbor, Michigan 48109, USA*

[1]*Department of Materials Science and Engineering, University of Michigan, Ann Arbor, Michigan 48109, USA*

InGaN light-emitting diodes (LEDs) are more efficient and cost effective than incandescent and fluorescent lighting, but lattice mismatch limits the thickness of InGaN layers that can be grown on GaN without performance-degrading dislocations. In this work, we apply hybrid density functional theory calculations to investigate the thermodynamic stability, lattice parameters, and band gaps of wurtzite and zincblende quaternary BInGaN alloys. We find that the wurtzite phase is more stable and can be lattice-matched to GaN for BInGaN compositions containing up to ~30% boron. The lattice match with GaN decreases strain and enables thicker active layers that mitigate Auger recombination and increase the efficiency of the LEDs. The band gap of the alloy remains tunable throughout the visible spectrum. Our results indicate that BInGaN alloys are promising alternatives to InGaN for high-efficiency, high-power LEDs.

## I. INTRODUCTION

InGaN light-emitting diodes (LEDs) enable significant energy and cost savings over incandescent and fluorescent lighting. Even greater cost savings can be achieved by increasing their efficiency or operating at higher power while keeping their efficiency high.[1] However, the ability of InGaN LEDs to operate with high efficiency at high power is hindered by the decrease of their internal quantum efficiency at high power (*efficiency droop*). This efficiency reduction is especially prevalent at longer wavelengths (*green gap*). Previous work has demonstrated that Auger recombination makes a significant contribution to the LED efficiency droop.[2] Since the Auger recombination rate is a function of a material-intrinsic Auger coefficient[3] and the electron and hole densities in the material, lowering the carrier density by increasing the InGaN quantum-well thickness and hence the active-region volume is the most promising and already demonstrated[4] method to reduce the Auger losses in InGaN. However, there are several materials challenges with this approach. The lattice mismatch between the InGaN active region and the underlying GaN layers causes performance-degrading dislocations to form if InGaN is grown beyond a certain critical thickness (e.g., 5 nm for $In_{0.15}Ga_{0.85}N$).[5] Furthermore, the currently used multiple-quantum-well (MQW) structures also do not solve the droop problem because for the same total active thickness, a single InGaN layer has higher efficiency than an MQW structure.

---

[a)] Author to whom correspondence should be addressed. Electronic mail: kioup@umich.edu.

This is because hole injection is poor through multiple layers in MQWs and only the first few QWs near the p-layer emit light.[6–8]

The lattice mismatch of InGaN to GaN can be addressed by co-alloying it with boron nitride to produce BInGaN alloys. These quaternary alloys maintain an approximate lattice match to GaN while allowing for a band gap that is adjustable throughout the visible range.[9] Co-alloying has previously been demonstrated as a method to independently vary the lattice constant and band gap in other material systems. Chai et al. found that the incorporation of Bi into $In_{0.53}Ga_{0.47}As$ to form $In_{0.53}Ga_{0.47}Bi_xAs_{1-x}$ increases the lattice constant and lowers the band gap for x up to 0.058.[10] Co-alloying has also been examined in $GaAs_{1-x-y}P_xBi_y$[11] and $GaAs_{1-x-y}N_xBi_y$[12,13], where the co-incorporation of P (N) and Bi yields alloys lattice-matched to GaAs and reduces the band gaps. Although BN preferentially crystallizes in the planar hexagonal structure, it is expected that for low BN content BInGaN alloys also adopt the wurtzite structure to minimize dangling bond formation and have a direct band gap determined primarily by the InGaN component. Previous work has shown that an approximate ratio of B:In ≈ 2:3 in wurtzite BInGaN produces alloys that are nearly lattice-matched to GaN.[9] This prediction agrees with a first-order approximation of the in-plane lattice constant of BInGaN by Vegard's law. Therefore, maintaining this B:In ratio while varying the Ga content enables the tuning of the band gap while still maintaining the lattice match to GaN that is necessary to increase the active region volume and achieve higher efficiencies.

These predictions are also supported by previous computational studies of BInGaN systems. Assali et al. used density functional theory (DFT) calculations with the local density approximation (LDA), generalized gradient approximation (GGA), and modified Becke-Johnson (mBJ) exchange-correlation functionals to examine nine different compositions of ordered cubic BInGaN. They found a lattice match at $B_{0.125}In_{0.187}Ga_{0.688}N$, which has approximately the 2:3 B:In ratio. Additionally, they found that In alloying reduces the band gap while B alloying raises the direct gap slightly and has a nonlinear effect on the (larger) indirect gap.[14] Park and Ahn examined wurtzite BInGaN/GaN quantum wells using multiband effective mass theory and report a lower mismatch to GaN than InGaN and a large reduction in the internal polarization field.[15] This reduction increases carrier wave-function overlap and ultimately the predicted efficiency of light production. However, it is necessary to expand on this existing work using more accurate computational methods and accouting for compositional disorder in random alloys.

In this work, we explore the thermodynamic, structural, and electronic properties of wurtzite and zincblende $B_yIn_xGa_{1-x-y}N$ alloys as a function of composition. We show that BInGaN alloys in both phases can be lattice-matched to GaN within their respective regions of thermodynamic stability by adjusting their composition. The wurtzite phase is more stable for alloys with lower boron content and can be lattice-matched to GaN for compositions up to ~30% boron. The wurtzite and zincblende phases of the alloy both have direct band gaps that are tunable throughout the visible the range. We also found that planar hexagonal BInGaN is not stable at any composition tested. Our results demonstrate that quaternary BInGaN alloys are promising materials to increase the high-power efficiency of nitride LEDs.

**II. METHODOLOGY**

We employed the Vienna *Ab initio* Simulation Package (VASP)[16–19] to perform DFT calculations using the projector augmented wave (PAW) method.[20,21] We used a 600 eV plane-wave cutoff and GW-compatible pseudopotentials that included 3, 13, 13, and 5 valence electrons for B, In, Ga, and N, respectively. For structural relaxations, we used the optB86b-vdW functional[22] and a Γ-centered Wisesa-McGill-Mueller Brillouin-zone grid with a minimum period distance of 21.48 Å.[23] We employed the optB86b-vdW functional for structural relaxation because van der Waals forces are relevant when analyzing the stability of BN and related compounds. Forces on atoms were relaxed to 1 meV/Å. We performed band-gap calculations with the Heyd-Scuseria-Ernzerhof (HSE06) functional.[24–26] We did not consider spin-orbit coupling effects on the band gap as they are weak in the nitrides (on the order of 10-20 meV)[27]: the conduction band is made of *s* orbitals for which the orbital angular momentum quantum number is zero and there is no spin-orbit coupling, while the valence band is made of nitrogen 2*p* orbitals, and since N is a light element spin-orbit coupling is weak. The publicly available Alloy Theoretic Automated Toolkit (ATAT)[28] and the Monte Carlo algorithm was used to model random alloys using Special Quasi-random Structures (SQSs). This was done using a 3x3x2 supercell for the wurtzite structures and 30-, 40-, and 48-atom supercells of variable shape for the zincblende and planar hexagonal structures. The arrangement of cations approximated the pair-correlation functions of random alloys up to 5.125 Å, which corresponds to the second-nearest neighbor in-plane and the nearest neighbor along the z-direction for wurtzite and the second-nearest neighbor in zincblende.

We tested several compositions at each of the 1:2, 2:3, 3:4, 4:5, and 1:1 B:In ratios. For each composition, we generated and relaxed an SQS in the wurtzite, zincblende, and hexagonal structures to determine structural parameters and thermodynamic properties. The Gibbs free energy was calculated using the analytical model for the entropy of a random solid solution, $S = -k_B \sum_{i=1}^{N} x_i \ln x_i$, where $x_i$ is the mole fraction for each of the N alloy ingredients, and $k_B$ is Boltzmann's constant. The hexagonal structures were not stable at any composition, so they were not included in future calculations. Since the 2:3, 3:4, and 4:5 ratios were closest to the expected lattice-matched compositions, we repeated the previous step four additional times at each composition for the wurtzite and zincblende structures (for a total of five SQSs at each composition). The electronic-structure calculations with the HSE06 functional were done for a single SQS for each composition that was selected because its band gap is close to the average PBE band gap of the five SQSs. We repeated these calculations for several alloys along the x- and y-axis of the $B_yIn_xGa_{1-x-y}N$ composition diagram (i.e., $B_yGa_{1-y}N$ and $In_xGa_{1-x}N$).

Finally, we fit models to the enthalpy of mixing, band gap, and lattice parameters as a function of composition for the wurtzite and zincblende structures. We used the regular solution model for the enthalpy of mixing,

$$\Delta H_{mix}(x,y) = xy\alpha_{BIn} + x(1-x-y)\alpha_{InGa} + y(1-x-y)\alpha_{BGa} \quad (1)$$

the bowing model for the band gap,

$$E_g(x,y) = E_g^{BN}y + E_g^{InN}x + E_g^{GaN}(1-x-y) + xy\beta_{BIn} + x(1-x-y)\beta_{InGa} + y(1-x-y)\beta_{BGa} \quad (2)$$

and a similar bowing model for the lattice parameters:

$$a(x,y) = a_{BN}y + a_{InN}x + a_{GaN}(1-x-y) + xy\delta_{BIn} + x(1-x-y)\delta_{InGa} + y(1-x-y)\delta_{BGa} \quad (3)$$

$$c(x,y) = c_{BN}y + c_{InN}x + c_{GaN}(1-x-y) + xy\theta_{BIn} + x(1-x-y)\theta_{InGa} + y(1-x-y)\theta_{BGa} \quad (4)$$

**III. RESULTS**

The thermodynamic, structural, and electronic properties of BInGaN alloys are plotted in Figs. 1 and 3-6. The colormap areas indicate the region that includes our explicit calculations at 1:2, 2:3, 3:4, 4:5, and 1:1 B:In composition ratios. The isolines illustrate how our fitted models (Eqs. 1-4) interpolate between the explicitly calculated BInGaN results and the data for ternary BGaN and InGaN alloys located on the axes.

**A. Thermodynamic Properties**

Our calculated results for the enthalpy of mixing and the phase-transition temperature between the solid solution and phase segregation are shown in Fig. 1, while our fitted model parameters to the enthalpy of mixing are listed in Tables I and II. Our analysis reveals that B-rich precipitates adopt the lowest-enthalpy planar hexagonal phase. However, small B-rich precipitates may adopt the zincblende crystal structure to minimize the interface energy with the embedding matrix[29], in agreement with our calculations. Both for the wurtzite and for the zincblende phases, the enthalpy of mixing and the phase-transition temperature increase as the boron mole fraction increases. Boron incorporation has a stronger effect on the enthalpy of mixing than indium. At indium mole fractions less than ~0.10 and boron mole fractions greater than ~0.03, indium addition reduces the phase-transition temperature due to increased configurational entropy. The phase-transition temperatures calculated here and in previous work[9] are well above the BInGaN growth temperatures used in experiment (typically around 700 °C).[30,31] This is because the phase-transition temperature represents thermodynamic equilibrium while growth of group-III nitrides is a non-equilibrium process. However, our calculated transition temperatures serve as a measure of the thermodynamic driving forces for phase separation and demonstrate that the quaternary BInGaN alloys are more stable against phase segregation than ternary alloys such as BGaN.[9]

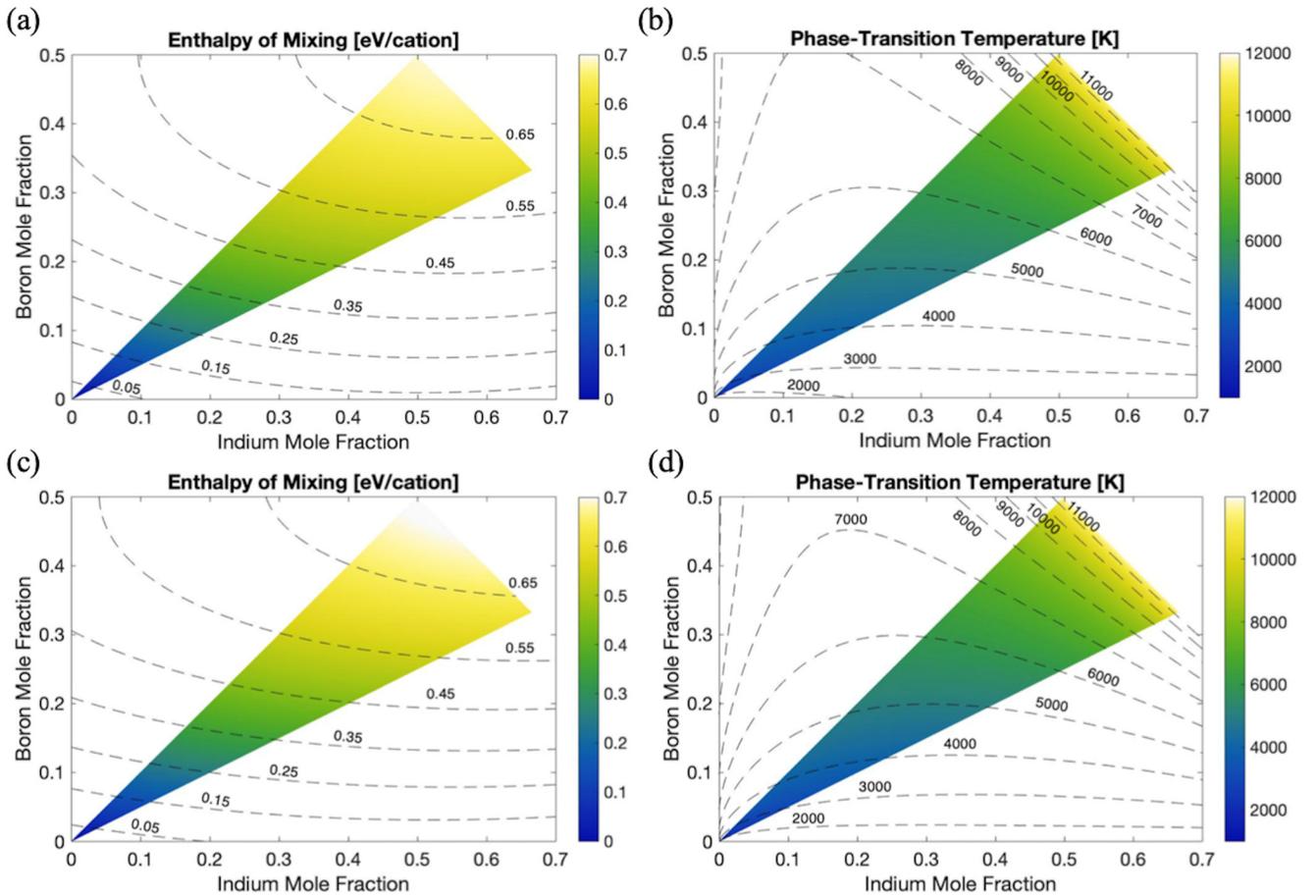

FIG. 1. Thermodynamics of (a, b) wurtzite and (c, d) zincblende $B_yIn_xGa_{1-x-y}N$ alloys. (a, c) Calculated enthalpy of mixing per cation and (b, d) phase-transition temperature as functions of composition. The enthalpy of mixing and the phase transition temperature increase as the boron mole fraction increases.

We also compare the relative stability of the wurtzite and the zincblende phases as a function of composition. The enthalpy of mixing for both phases was calculated with respect to the most stable phase of each binary compound (wurtzite for InN and GaN, and hexagonal for BN). Our results (Fig. 2) show that the wurtzite phase is more stable for quaternary BInGaN compositions closer to ternary InGaN, while the zincblende phase is more stable for those closer to ternary BGaN. For B compositions up to ~30%, the wurtzite phase of $B_yIn_xGa_{1-x-y}N$ can be stabilized with the addition of In and can even be lattice matched to GaN by the co-incorporation of B and In at the appropriate ratio (discussed in Section III.B). This agrees with the expectation that the alloy adopts the wurtzite structure at low BN content to minimize dangling-bond formation. At boron compositions higher than 30%, the zincblende phase is the most stable independent of In content. The zincblende-to-wurtzite transition curve in Fig. 2 is independent of temperature because the configurational entropy per cation (and hence the temperature dependence of the free energy) is the same for both structures. Our approach does not account for other sources of entropy (vibrational, electronic, defect, etc.) that may differ for the two structures. However, configurational disorder is the dominant entropy term for the non-dilute alloy compositions that occur at the transition curve.

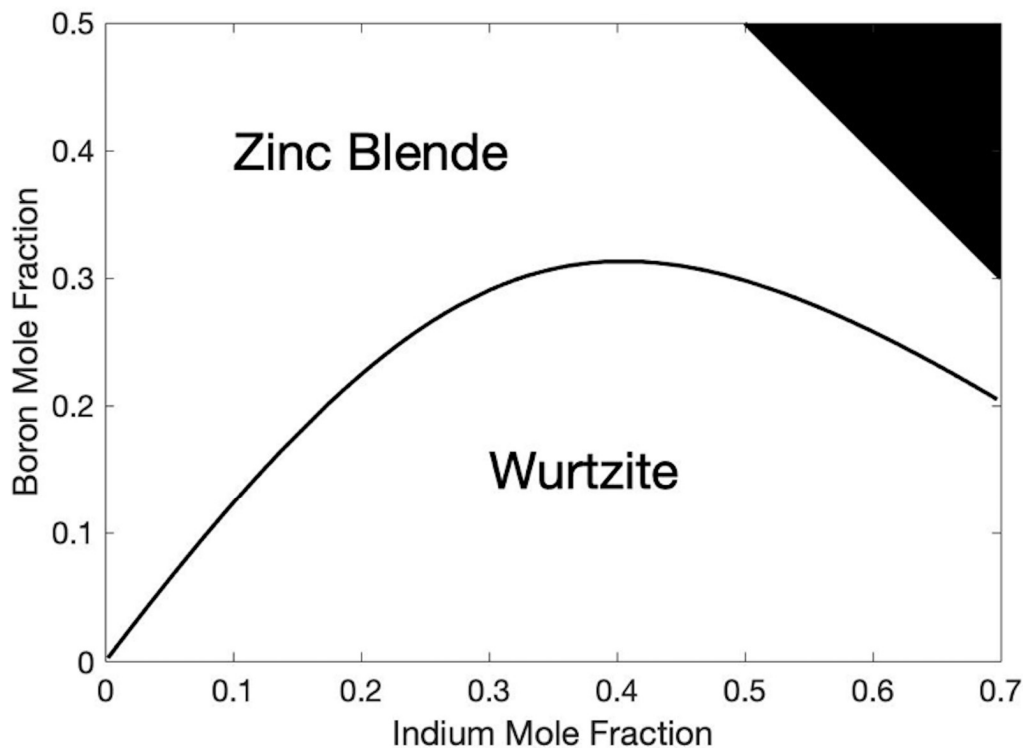

FIG. 2. Relative thermodynamic stability of wurtzite and zincblende $B_yIn_xGa_{1-x-y}N$ alloys. The wurtzite phase is more stable at compositions closer to InGaN, and the zincblende phase is more stable at compositions closer to BGaN.

We also found that the planar hexagonal structure is less stable than both the wurtzite and zincblende structures at all compositions tested. We performed structural relaxations on planar hexagonal structures throughout the composition range. Of the structures tested, most relaxed to a tetrahedral form (zincblende, wurtzite, or amorphous). The structures that remained planar hexagonal were generally higher in energy than the zincblende structure at the same composition by about 0.5 eV per cation.

**B. Structural Parameters**

We subsequently examined the lattice constants of the wurtzite and zincblende phases of $B_yIn_xGa_{1-x-y}N$ as a function of composition. Our results are plotted in Figs. 3 and 4, and our fitted model parameters are listed in Tables I and II. For compositions up to ~5% B in the wurtzite phase, maintaining a composition ratio of approximately 2 parts B to 3 parts In provides a lattice match to GaN along the *a* direction (Fig. 3). The lattice-matched ratio increases at higher B compositions, reaching as high as ~3:2 for B compositions above 40%. This may be a result of the large size mismatch between B and Ga/In, since atomic-size differences have been shown to contribute to deviations from Vegard's law.[32] The zincblende structure can also be lattice-matched to GaN, but the B to In ratio required for matching is much lower than that required for wurtzite (approximately 1 part B to 7 parts In at 5% B composition). The zincblende structural data is illustrated in Fig. 4.

By setting the left-hand side of Eqn. 3 equal to the *a* lattice parameter of wurtzite GaN, we can solve for the B composition (y) that provides a lattice match to GaN for a given In composition (x):

$$y = 2.357\left(\sqrt{x^2 - 1.455x + 1.049} + 0.984(x - 1.041)\right), \tag{5}$$

and similarly, the composition for matching the *c* lattice parameter in wurtzite is:

$$y = 4.124 \times 10^{-8}\left(\sqrt{-2.935 \times 10^{14}(x^2 - 57.986x - 252.834)} - 1.619 \times 10^7(x + 16.822)\right). \tag{6}$$

For the *a* lattice parameter in zincblende the B:In ratio lattice-matched to GaN is given by:

$$y = -3.966\left(\sqrt{x^2 - 1.342x + 0.387} + 1.050(x - 0.591)\right). \tag{7}$$

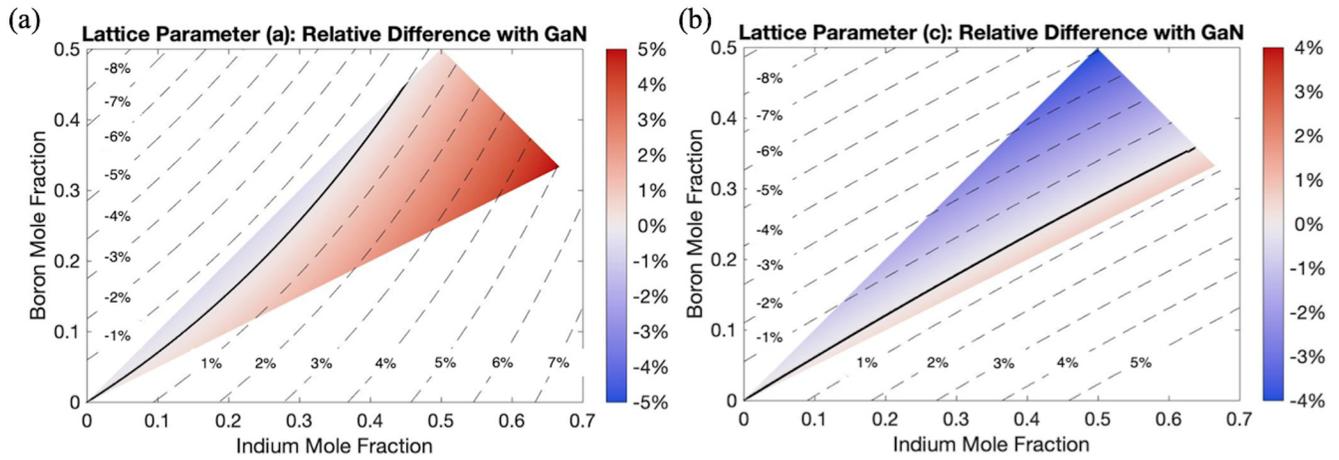

FIG. 3. Relative difference of wurtzite $B_yIn_xGa_{1-x-y}N$ calculated lattice constants with that of GaN as a function of composition along the (a) $a$ and (b) $c$ directions. The solid black lines indicate an exact lattice match with GaN along each direction. The lattice-match line in (a) corresponds to Eqn. 5, and that in (b) corresponds to Eqn. 6. A ratio of approximately 2 B to 3 In provides a lattice match to GaN in the $a$ direction for up to ~5% B composition.

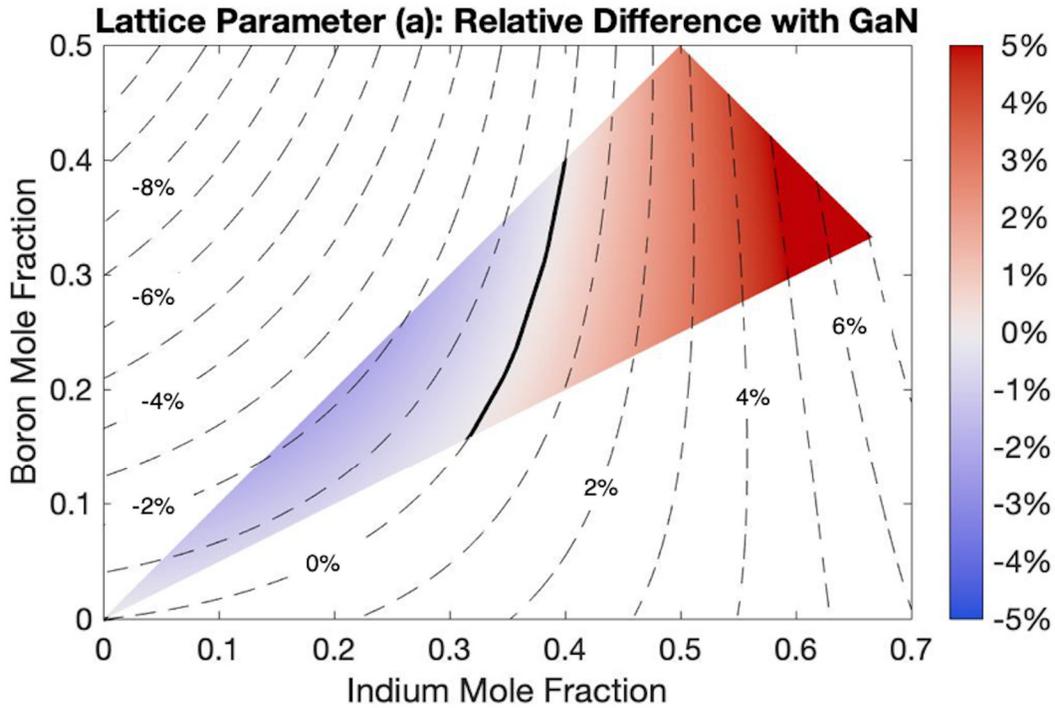

FIG. 4. Relative difference of zincblende $B_yIn_xGa_{1-x-y}N$ calculated lattice constant with that of GaN as a function of composition. The solid black line indicates an exact lattice match with GaN and corresponds to Eqn. 7.

## C. Band Gaps

We also calculated the band gap of $B_yIn_xGa_{1-x-y}N$ alloys and found that they span the entire visible spectrum both for the wurtzite and for the zincblende phases (Figs. 5 and 6). The fitted parameters to the gap-bowing model are listed in Tables I and II. The gap values calculated with HSE06 have been rigidly shifted by 0.25 eV (wurtzite) and 0.31 eV (zincblende) to match the experimental gaps (3.39 eV and 3.24 eV, respectively) of GaN at room temperature.[33–35] For both phases, the band gap monotonically decreases as the indium content increases, while the band gap displays strong bowing parameters with respect to B-In and B-Ga. Our calculated In-Ga bowing parameters (1.38 eV) is in good agreement with the value of Moses and Van de Walle (1.36 eV).[36] The B-In and In-Ga bowing parameters (listed in Tables I and II) are larger in magnitude for wurtzite, while the B-Ga bowing parameter is larger in magnitude for zincblende. In the wurtzite phase, small additions of boron to high-indium-content InGaN lower the band gap due to the large B-In bowing parameter. For zincblende, increasing boron mole fractions up to ~0.15 decreases the band gap, while increasing the boron amount beyond ~0.15 acts to increase the gap.

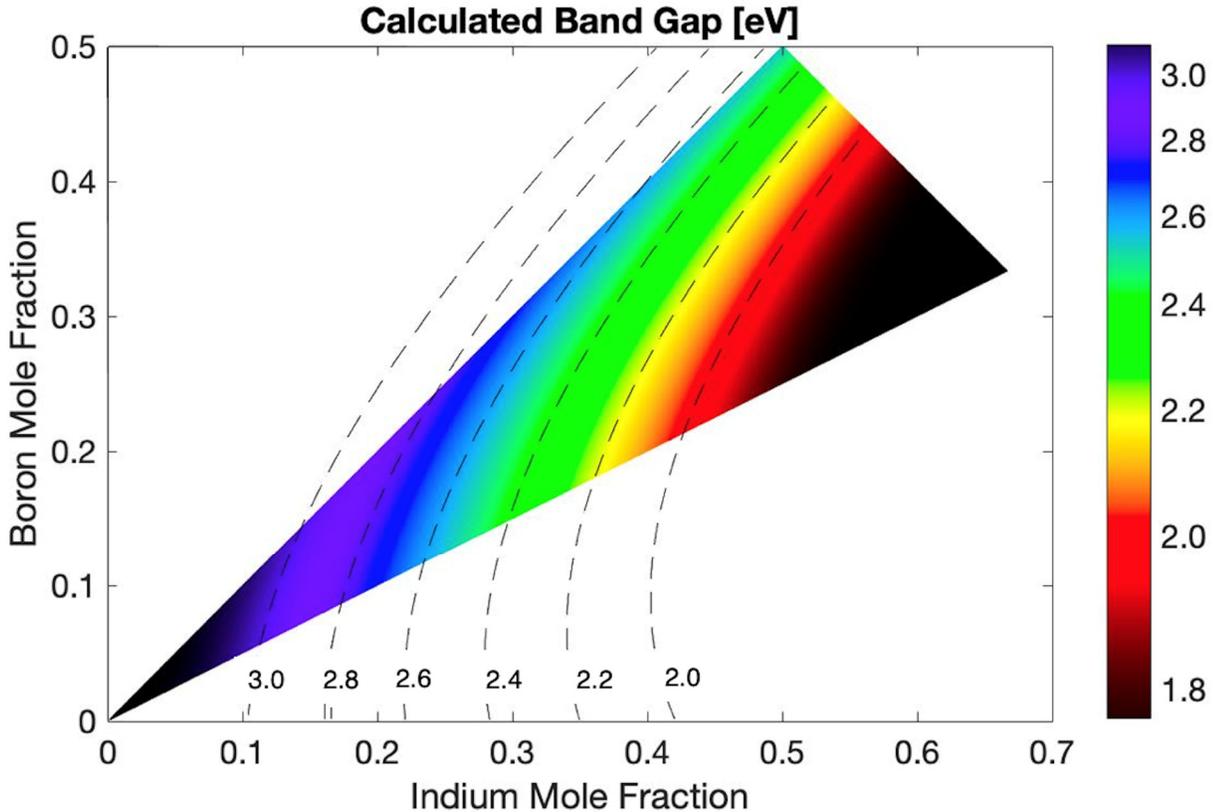

FIG. 5. Calculated band gap of wurtzite $B_yIn_xGa_{1-x-y}N$ as a function of boron and indium mole fractions, depicted by the corresponding color on the visible light spectrum. The gap values calculated with HSE06 have been rigidly increased by 0.25 eV to match the experimental gap of wurtzite GaN at room temperature (3.39 eV).[33]

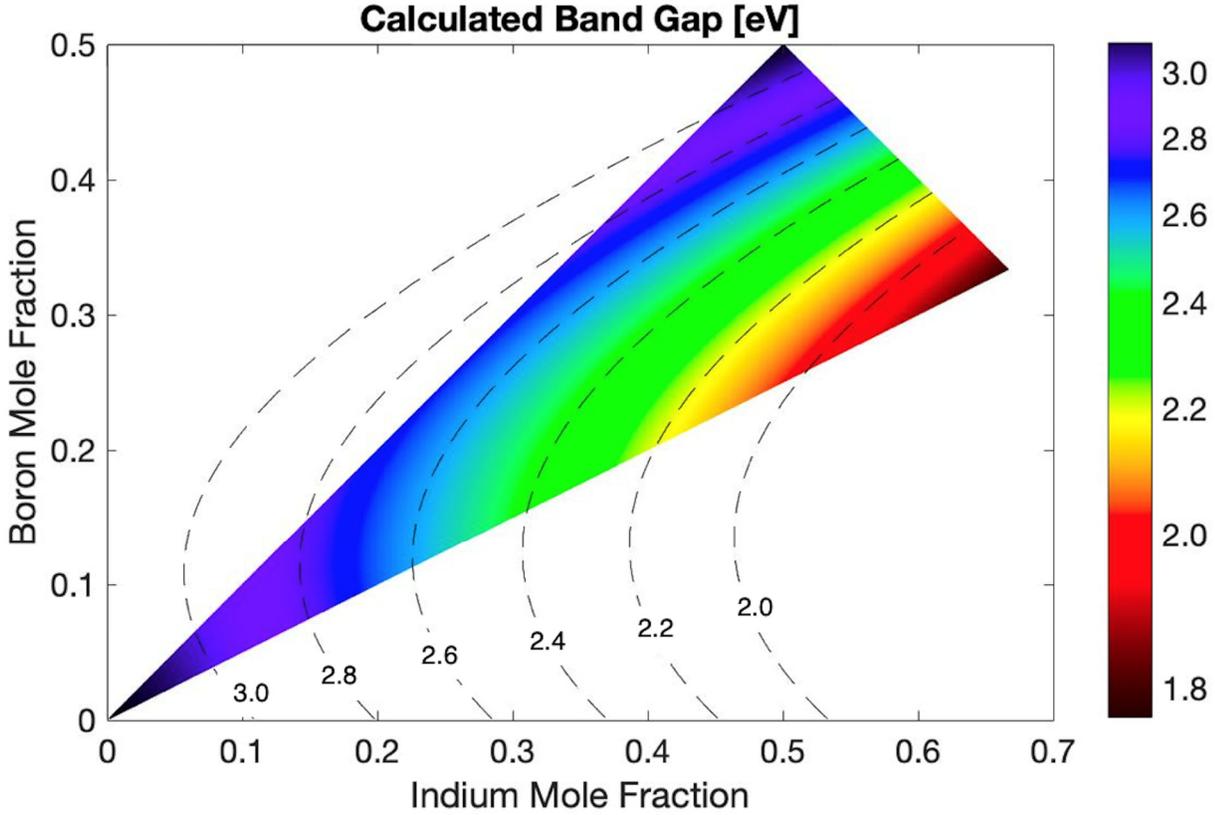

FIG. 6. Calculated band gap of zincblende $B_yIn_xGa_{1-x-y}N$ as a function of boron and indium mole fractions, depicted by the corresponding color on the visible light spectrum. The gap values calculated with HSE06 have been rigidly increased by 0.31 eV to match the experimental gap of zincblende GaN at room temperature (3.24 eV).[34,35]

**D. Lattice-Matched Compositions**

The fitting parameters for each model are shown in Tables 1 and 2, and the data for wurtzite BInGaN lattice-matched to GaN along the *a* direction are summarized in Fig. 7. When limited to the lattice-matched compositions represented by Eqn. 5, the phase-transition temperature and enthalpy of mixing per cation monotonically increase as the In mole fraction increases. The enthalpy of mixing increases approximately linearly up to about 35% In. The band gap decreases up to about 38% In and subsequently increases for higher In content.

TABLE I. Fitting parameters for the enthalpy of mixing, band gap, and lattice constant of wurtzite BInGaN alloys as a function of composition (Eqs. 1-4).

| Property | $\Delta H_{mix}$ | $E_g$ | $a$ | $c$ |
|---|---|---|---|---|
| Parameter | $\alpha$ | $\beta$ | $\delta$ | $\theta$ |
| B-In | 2.76 ± 0.04 | -12.25 ± 0.23 | 0.60 ± 0.02 | 0.03 ± 0.05 |
| In-Ga | 0.52 ± 0.06 | -1.38 ± 0.20 | -0.02 ± 0.04 | 0.04 ± 0.08 |
| B-Ga | 1.97 ± 0.08 | -6.15 ± 0.25 | 0.11 ± 0.04 | 0.04 ± 0.10 |

TABLE II. Fitting parameters for the enthalpy of mixing, band gap, and lattice constant of zincblende BInGaN alloys as a function of composition (Eqs. 1-3).

| Property | $\Delta H_{mix}$ | $E_g$ | $a$ |
|---|---|---|---|
| Parameter | $\alpha$ | $\beta$ | $\delta$ |
| B-In | 2.84 ± 0.05 | -10.50 ± 1.47 | 1.30 ± 0.13 |
| In-Ga | 0.32 ± 0.04 | 0.35 ± 0.82 | -0.37 ± 0.11 |
| B-Ga | 2.12 ± 0.06 | -9.66 ± 0.93 | -0.23 ± 0.14 |

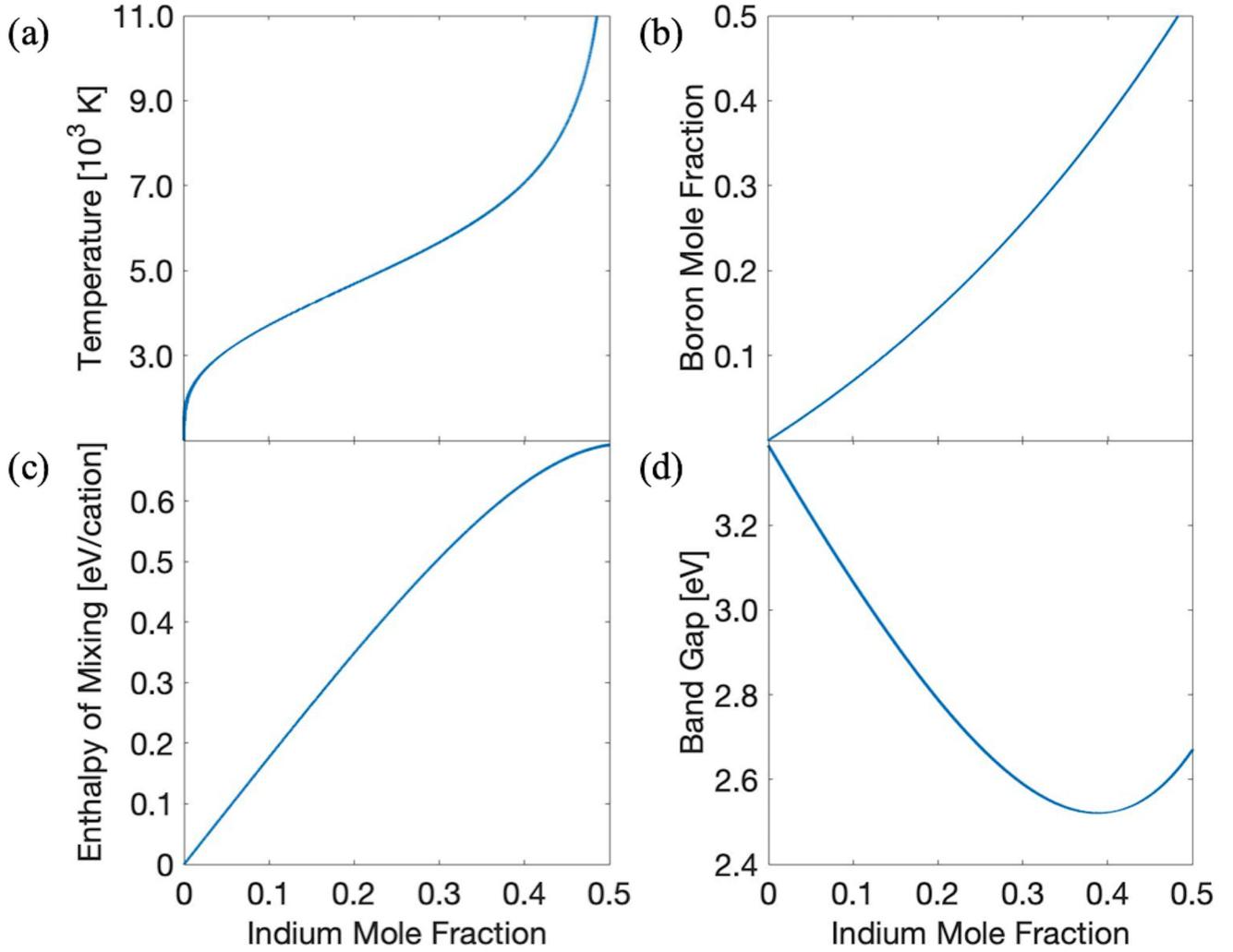

FIG. 7. Calculated values for (a) phase-transition temperature, (b) enthalpy of mixing per cation, (c) boron mole fraction, and (d) band gap for wurtzite $B_{y(x)}In_xGa_{1-x-y(x)}N$ compositions exactly lattice matched to GaN along the *a* direction. As in Figure 6, the HSE06 calculated gap values have been rigidly increased by 0.25 eV to match the experimental gap of GaN at room temperature (3.39 eV).[33] The lattice-matched curve in (b) is represented by Eqn. 5.

**E. Discussion**

Our work using hybrid functionals agrees with that of Assali et al. using mBJ potentials to demonstrate that co-incorporation of B and In alloying reduce the band gap of zincblende GaN.[14] However, our results show that although the addition of boron content up to ~0.15 decreases the band gap, continuing to increase it beyond ~0.15 acts to increase the gap, while their work found a monotonic increase in the direct band gap with increasing boron addition.

Although our work has developed a more thorough understanding of the properties of BInGaN alloys, the validation of our predictions by growth experiments is challenging. Only the growth of low-boron-content quaternary BInGaN alloys has been achieved currently by Gautier et al.[30,31] Gautier et al. reported smaller gaps and lattice constants than GaN for up to 2% B and up to 14% In composition for 140 nm thick layers of BInGaN grown on GaN/sapphire and ZnO-buffered Si substrates with metal-organic vapor phase epitaxy (MOVPE).[30,31] Our results are in excellent agreement with their experimental measurements (Table III). However, our calculations predict that higher B content alloys may be possible to realize experimentally. We can predict an upper limit on B content (using existing growth approaches) based on the demonstrated or predicted composition limits of $B_yGa_{1-y}N$. It is estimated that 1.5 times as much B can be incorporated into InGaN compared to GaN.[9] Cramer et al. observed up to 3% B incorporation into high-crystal-quality BGaN using plasma assisted molecular beam epitaxy (MBE).[37] This would lead to ~4.5% B incorporation in BInGaN. Calculations by Lymperakis predict that up to 25% B incorporation should be possible using MBE under N-rich conditions.[38] This growth technique presents its own challenges (e.g., lower crystal quality when grown under N-rich conditions,[38] different temperatures needed for growth of boron, gallium, and indium, and the higher cost of MBE compared to alternative techniques), but indicates that there may be alternate growth conditions that allow for far higher boron incorporation than has currently been achieved in group-III nitride growths to date.

TABLE III. Comparison of theoretical values for out-of-plane lattice constant ($c$) and band gap ($E_g$) of BInGaN to available experimental data.[30]

| Composition | $c$ (Ref. [30]) [Å] | $c$ (This Work) [Å] | $E_g$ (Ref. [30]) [eV] | $E_g$ (This Work) [eV] |
|---|---|---|---|---|
| $In_{0.195}Ga_{0.840}N$ | 5.29 | 5.32 | 2.53 | 2.43 |
| $B_{0.010}In_{0.160}Ga_{0.830}N$ | 5.25 | 5.29 | 2.58 | 2.55 |
| $B_{0.020}In_{0.140}Ga_{0.805}N$ | 5.23 | 5.23 | 2.66 | 2.63 |

## IV. CONCLUSION

In summary, we applied hybrid DFT calculations to explore the properties of quaternary BInGaN alloys. Our results are in agreement with previous work that predicts a lattice match of BInGaN to GaN along the *a* direction with an approximately 2:3 ratio of B:In content. We have compared the thermodynamic stability of the wurtzite, zincblende, and planar hexagonal phases of BInGaN and determined the regions of relative stability. Our analysis shows that the wurtzite phase is more stable and can be lattice matched to GaN for BInGaN compositions containing up to ~30% boron. As a result, co-alloying InGaN with BN produce wurtzite BInGaN alloys that maintain an approximate lattice match to GaN while allowing for a band gap that is adjustable throughout the visible range. This lattice match decreases strain, which in turn allows for the growth of thicker active layers without performance-degrading dislocations. The thicker active layer decreases the carrier concentration at a given carrier density and ultimately mitigates Auger recombination. Therefore, wurtzite BInGaN alloys offer a solution to the efficiency-droop and green-gap problems of InGaN LEDs and are promising material for higher efficiency high-power visible LEDs.

**ACKNOWLEDGMENTS**

This work was supported by the Designing Materials to Revolutionize and Engineer our Future (DMREF) Program under Award No. 1534221, funded by the National Science Foundation. It used resources of the National Energy Research Scientific Computing Center (NERSC), a U.S. Department of Energy Office of Science User Facility operated under Contract No. DE-AC02-05CH11231.

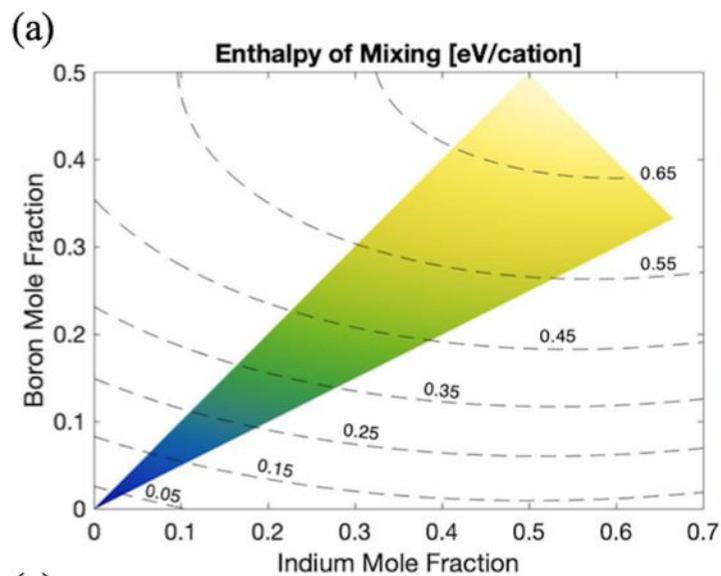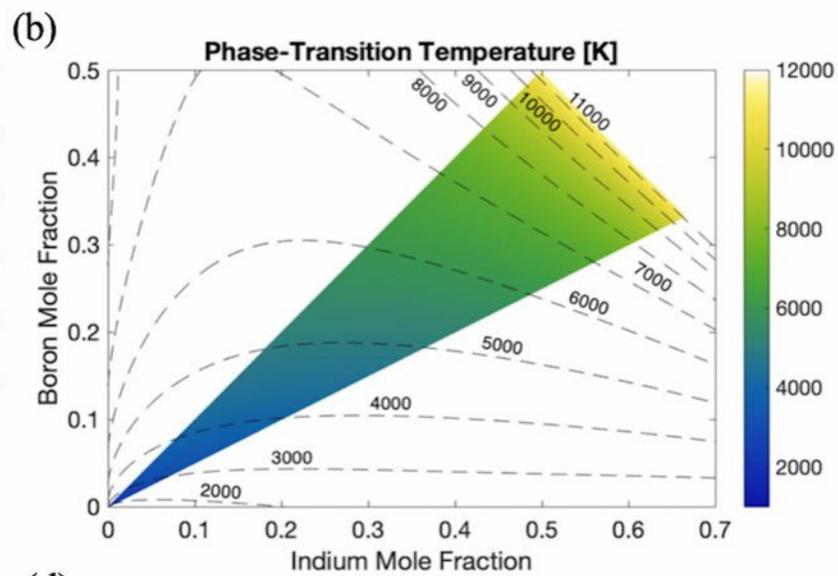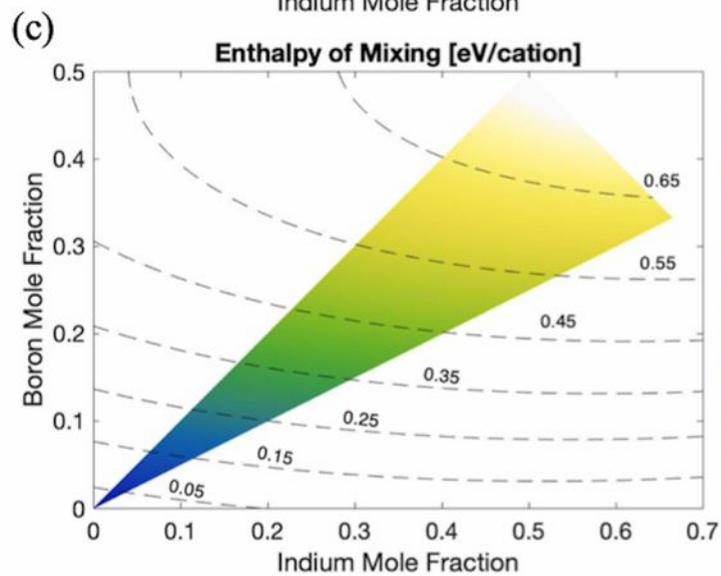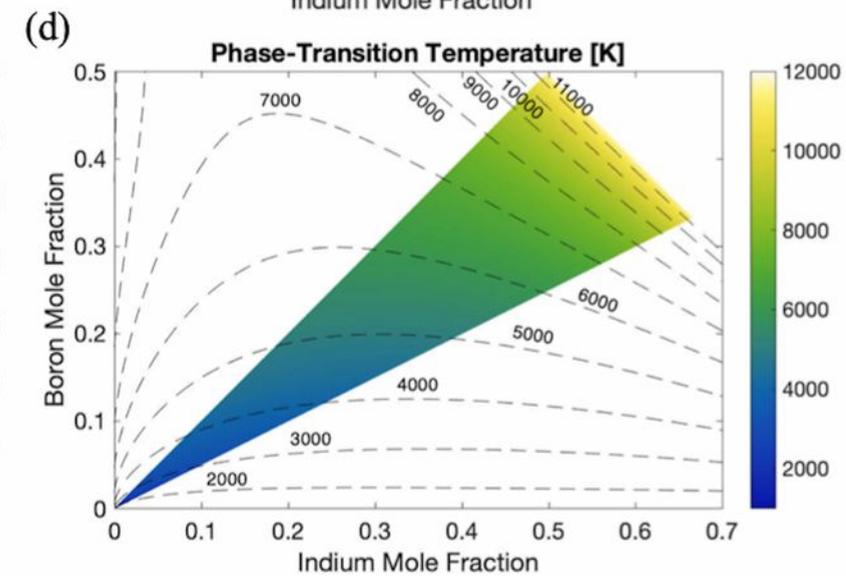

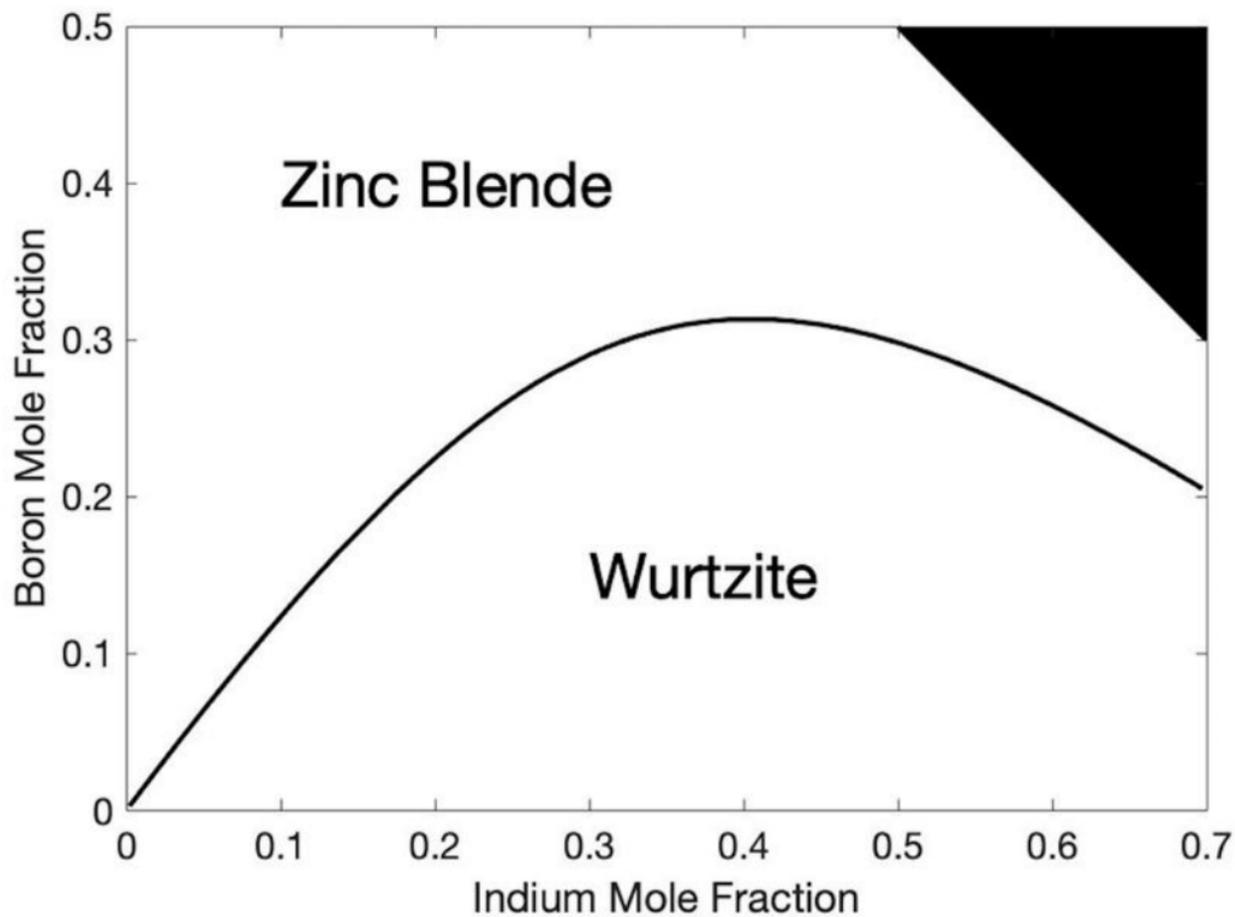

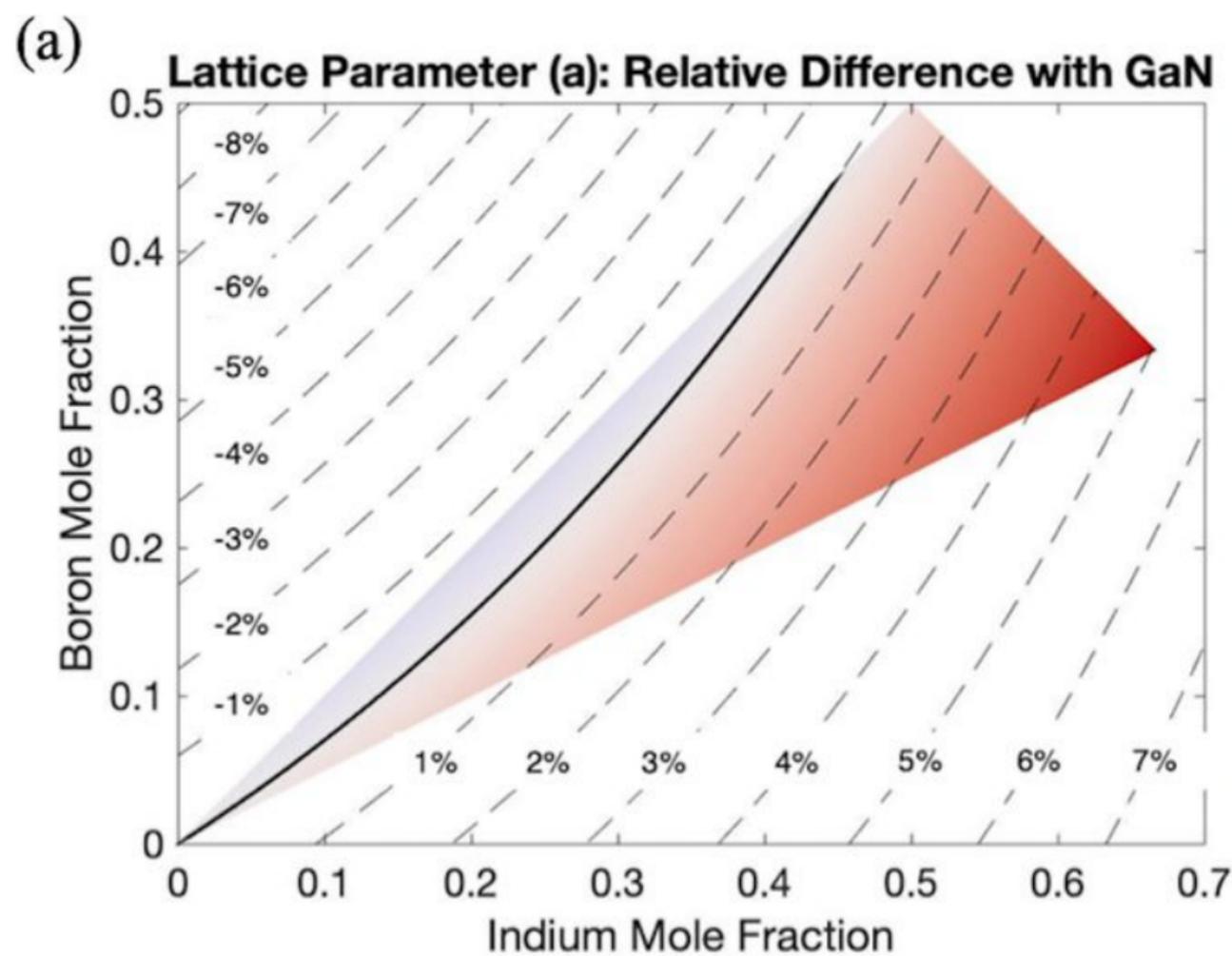 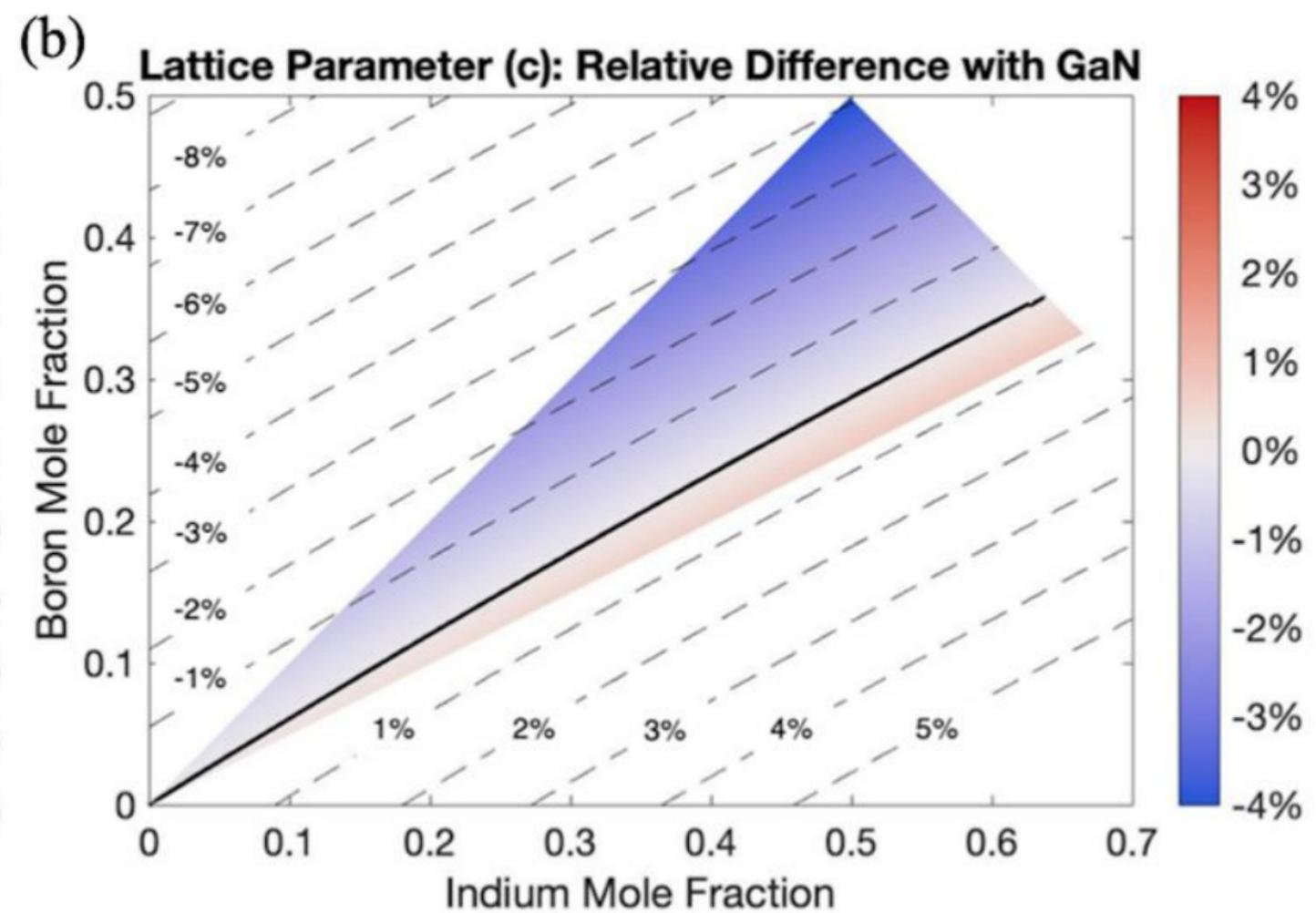

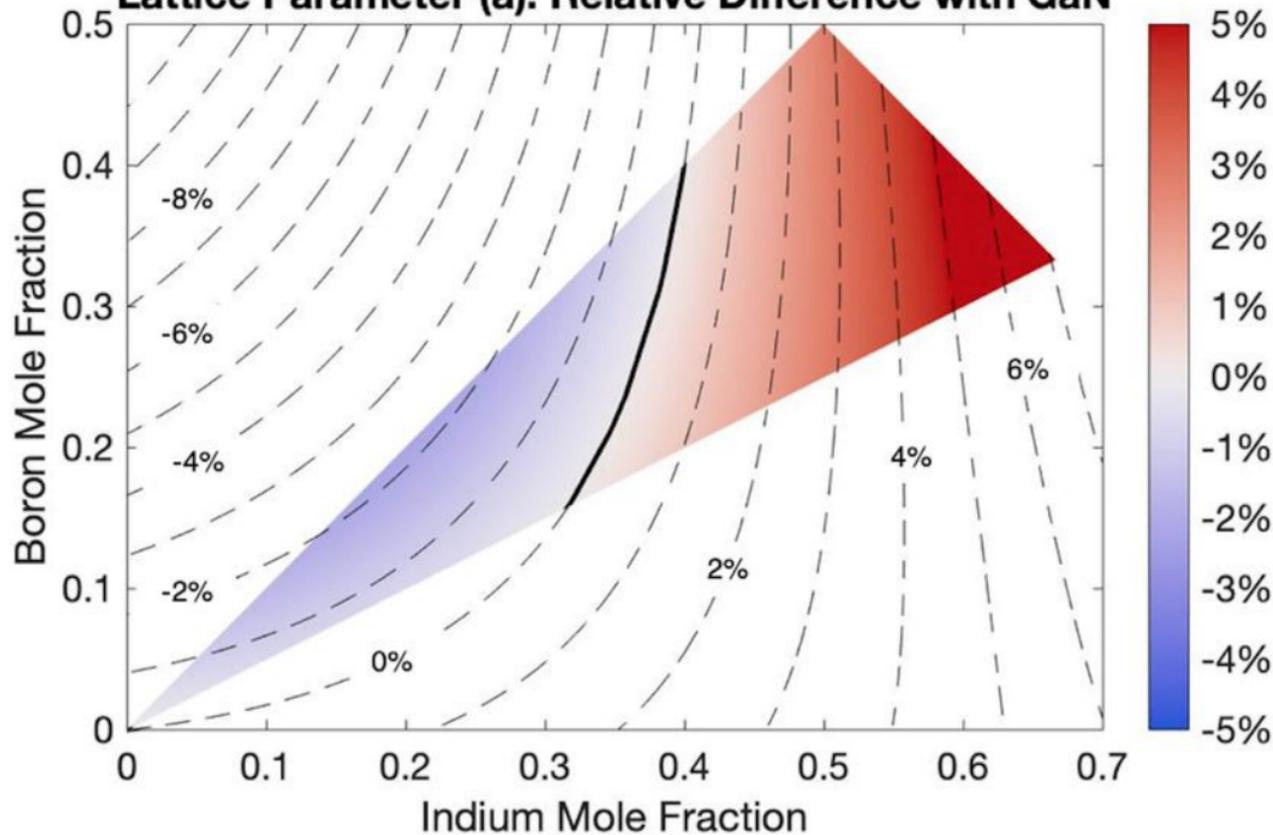

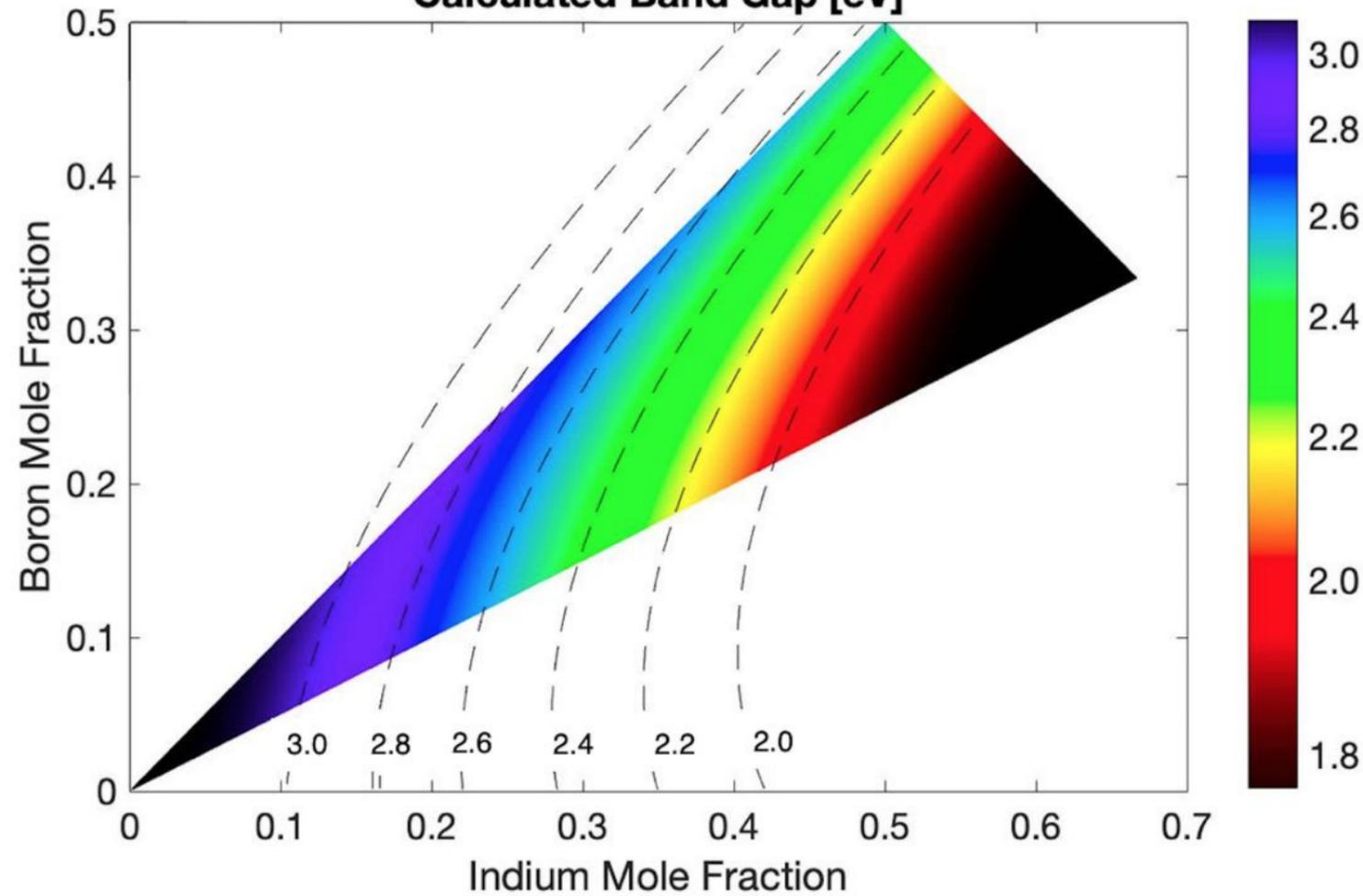

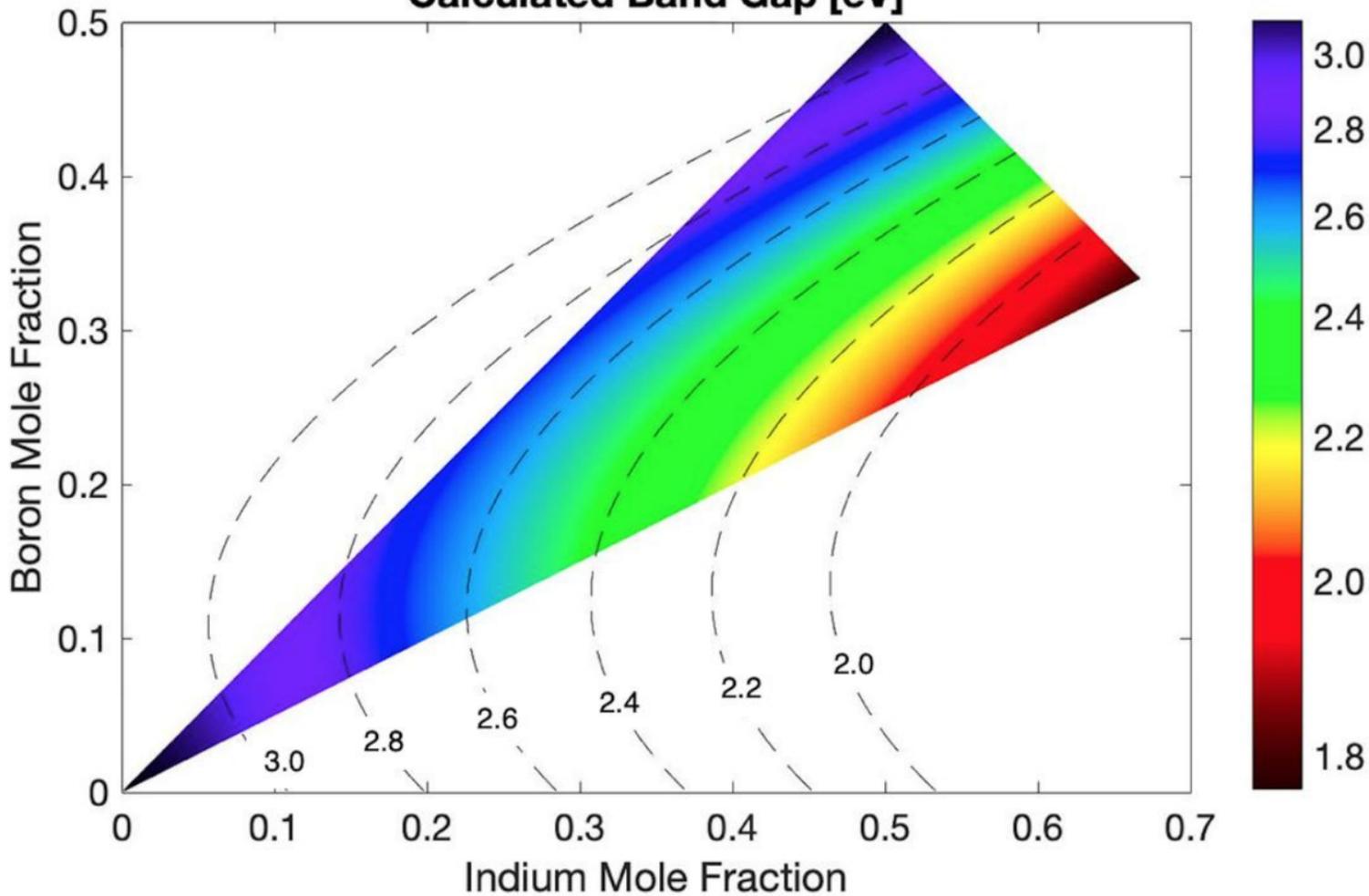

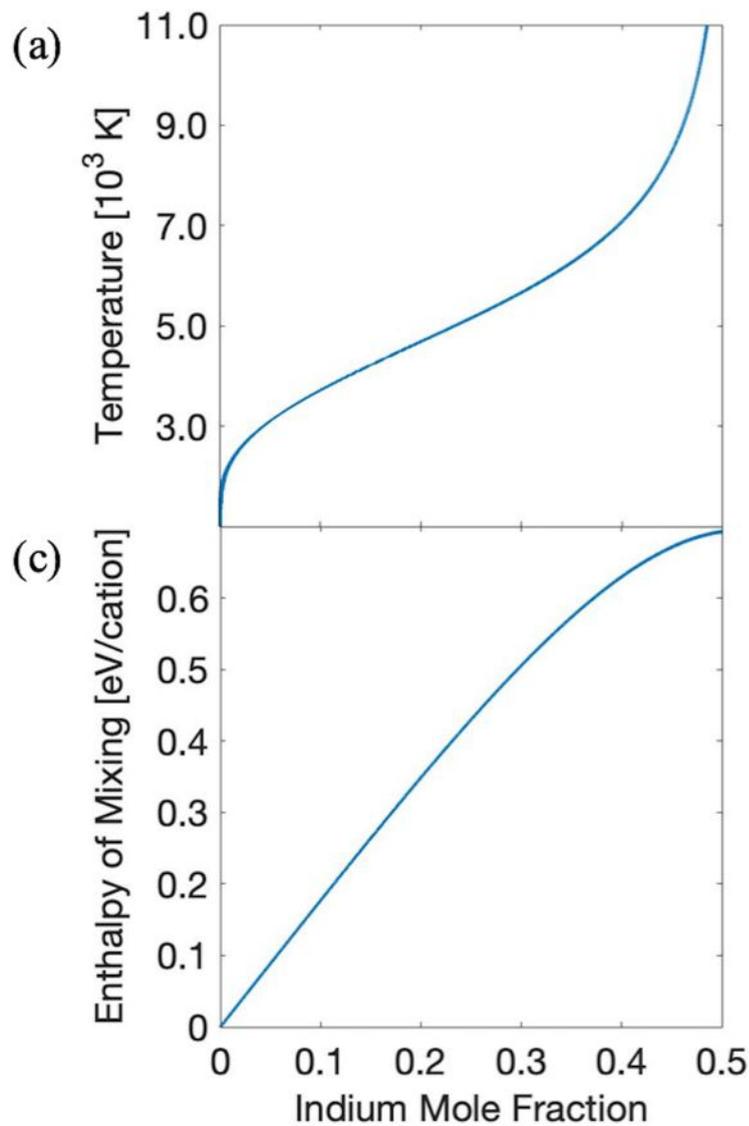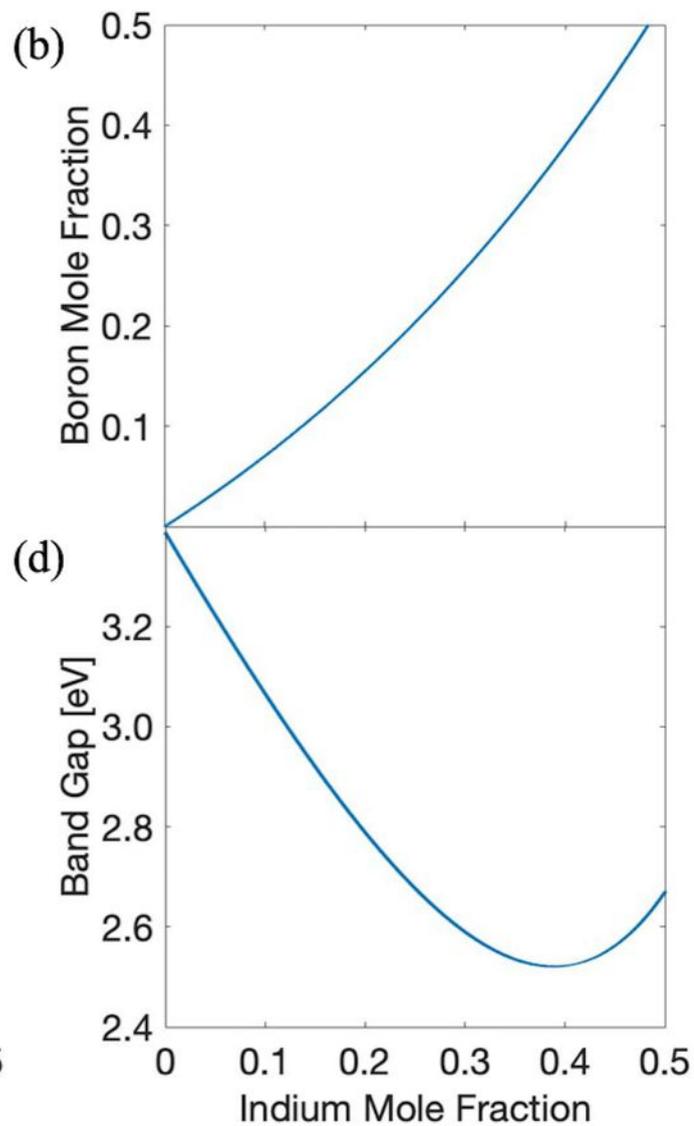